\documentclass[aps,prb,twocolumn,amsmath,amssymb,nofootinbib,superscriptaddress,floatfix]{revtex4-1}

\usepackage{color}
\usepackage{xparse}
\usepackage{amsmath}
\usepackage{physics}
\usepackage{braket}
\usepackage{graphicx}
\usepackage{lipsum}
\usepackage{csquotes,helvet,mathpazo,listings}
\usepackage{notes2bib}
\usepackage{dcolumn}
\usepackage{bm}
\usepackage[normalem]{ulem}
\usepackage{verbatim}

\newcommand{\bs}[1]{{\boldsymbol{#1}}}

\begin{document}

\title{
Chern Bands with Higher-Order Van Hove Singularities on Topological Moir\'e Surface States
}
\author{Lakshmi Pullasseri}
\affiliation
{
Department  of  Physics,  Emory  University,  400 Dowman Drive, Atlanta,  GA  30322,  USA
}
\author{Luiz H. Santos}
\affiliation
{
Department  of  Physics,  Emory  University,  400 Dowman Drive, Atlanta,  GA  30322,  USA
}

\date{\today}

\begin{abstract} 
In two-dimensional electronic lattices, changes in the topology of the Fermi surface (Lifshitz transitions) lead to Van Hove singularities characterized by a divergence in the electronic density of states. {Van Hove singularities can enhance the effect of electronic interactions, providing a platform to explore novel correlated electronic states.}
In this work, we investigate the emergence of topological Chern bands on the surface of three-dimensional topological insulators, which host higher-order Van Hove singularities that are characterized by the power-law diverging density of states. These singularities can arise from the interplay between a time-reversal breaking Zeeman field induced by proximity to a ferromagnetic insulator and a time-reversal invariant moiré potential on the surface electrons, created by quintuple layer misalignment in a family of topological insulators such as Bi$_2$Se$_3$ and Bi$_2$Te$_3$, which host a single surface Dirac fermion. 
We establish the onset of Chern bands near charge neutrality with Chern numbers $C = \pm 1$ that also possess a manifold of higher-order Van Hove singularities on the moiré Brillouin zone valleys controlled by the Zeeman and moir\'e potential energy scales, unveiling a new platform to realize exotic Lifshitz transitions in topological bands.
Furthermore, we show that the strong peaks in the density of states in the vicinity of Lifshitz transitions give rise to characteristic features in the low-temperature intrinsic anomalous Hall conductivity, yielding a path to probe Van Hove singularities in Chern bands through anomalous transport measurements.
\end{abstract}

\maketitle


\section{Introduction}

In two-dimensional systems, topological changes of the Fermi surface, known as Lifshitz transitions \cite{lifshitz1960anomalies}, are marked by saddle points in the energy dispersion, which give rise to Van Hove singularities (VHS) where the density of states has a logarithmic divergence. \cite{VanHove1953TheOccurrence} 
Van Hove has demonstrated that these singularities are generic features of two-dimensional periodic systems, and the existence of saddle points in the periodic energy dispersion $\varepsilon(\bs{k})$ is of topological origin and can be understood within the framework of Morse theory.
{Recent progress in the synthesis of new classes of two-dimensional materials along with the possibility of engineering electronic band structures have renewed interest in characterizing electronic states near VHS.}
In particular, the enhancement of interaction effects due to the large accumulation of electronic states in the vicinity of VHSs has been studied as a fruitful setting to characterize Fermi liquid instabilities in cuprates \cite{schulz1987superconductivity,dzyaloshinskiui1987maximal,markiewicz1997survey},
doped graphene \cite{Nandkishore2012,WangFunctionalRG2012,KieselCompeting2012,GonzalezGraphene2008}, 
Hofstadter systems \cite{shaffer2022unconventional,shaffer2023triplet}
and moir\'e graphene superlattices
\cite{wu_chern_2021,isobe2018unconventional, Sherkunov-Betouras2018, Liu-Chiral2018, Kennes-Strongcorrelations-2018, You-Vishwanath-2019, Lin-Nandkishore-2020, hsu_topological_2020, classen_competing_2020, Chichinadze2020Nematicsuperconductivity}.

A physically rich scenario occurs when the energy dispersion of electronic bands supports higher-order saddles, giving rise to higher-order Van Hove singularities (HOVHS) characterized by stronger power-law divergence of the density of states.\cite{shtyk2017electrons,yuan2019magic}
While previous studies of HOVHS have mainly focused on the properties of time-reversal invariant bands, the interplay between HOVHS and time-reversal broken Chern bands has been recently highlighted as a path towards new electronic phases. \cite{castro2023emergence,aksoy2023single} 
In particular,
Ref. \onlinecite{castro2023emergence} has mapped the landscape of VHS of Haldane Chern bands on the honeycomb lattice \cite{Haldane1988} and demonstrated that, under inversion symmetry, the system supports a pair of HOVHS at the two valleys of the Brillouin zone which, under repulsive interactions, give rise to a rich phase diagram containing pair density wave superconductivity and a Chern super metal state. These developments highlight the importance of identifying new electronic platforms supporting Chern bands with HOVHS. 

In this work, we provide a route to realize Chern bands supporting HOVHS on the surface of three-dimensional topological insulators (TI). \cite{moore2010birth,hasan2010colloquium,qi2011topological} Our mechanism combines the breaking of time-reversal symmetry on the TI surface via proximity to a ferromagnetic insulator and the presence of a moir\'e pattern.
Remarkably, in topological insulators hosting a single Dirac fermion on the surface, such as Bi$_2$Te$_3$ and Bi$_2$Se$_3$, a moir\'e pattern can emerge when the crystals are grown using the technique of molecular beam epitaxy (MBE). In the case of Bi$_2$Te$_3$\cite{chen2009experimental} a small in-plane rotation of the top layer (facilitated by Cu dopants that reduce the interlayer coupling) results in the emergence of a triangular moir\'e superlattice of constant $\approx 13 nm$. 
\cite{schouteden2016moire} 
Moreover, on Bi$_2$Se$_3$ 
\cite{kuroda2010hexagonally} 
grown using MBE on a substrate, superstructures with lattice constants $\approx 10 nm$ can be constructed from the direct lattice mismatch of the crystal with substrates like graphene, \cite{song2010topological} FeSe,\cite{wang2012scanning} Au$(111)$,\cite{jeon2011metal} hBN\cite{xu2015van} and In$_2$Se$_3$. \cite{wang2011superlattices}
The interplay of Dirac surface states with a time reversal invariant moir\'e surface potential reorganizes the surface states into a sequence of moir\'e bands hosting satellite Dirac fermions.\cite{wang2021moire,cano2021moire}. Furthermore, near charge neutrality, one of these time-reversal invariant bands can support a pair of HOVHS with cubic dispersion on each of the valley points $\pm \bs{K}$ of the moir\'e Brillouin zone.\cite{wang2021moire}. 

We explore the effects of a Zeeman field on these HOVHS states near charge neutrality. The addition of a Zeeman field significantly modifies the characters of the HOVHS band near charge neutrality. First, the Zeeman field gaps out this band resulting in a pair of topological bands with Chern numbers $\pm 1$ near charge neutrality. Second, the Zeeman field behaves as a control knob for the character of the Lifshtz transition. Specifically, while the time-reversal symmetric pair of HOVHS occurs for a fine-tuned value of the moir\'e potential $V$,\cite{wang2021moire} the Zeeman field gives rise to a line HOVHS in the $(V,h)$ plane. Remarkably, we uncover simple expressions for these lines of HOVHS in a Chern band for moir\'e potentials with C$_3$ and C$_6$ rotation symmetries, and we estimate that this HOVHS manifold could be within experimental reach for certain TI systems. These results establish a promising route to achieve higher-order Lifshitz transitions in time-reversal broken TI surface states. 

The existence of exotic Lifshtz transitions in topological Chern bands raises the prospect of exploring new interplays between Van Hove singularities and band topology. Along these lines, we establish a connection between Lifshitz transitions and the quantum geometry of Chern bands through the intrinsic part of the anomalous Hall conductivity $\sigma^{\textrm{int}}_{xy}$. \cite{karplus1954hall,jungwirth2002anomalous,nagaosa2010anomalous}
The intrinsic Hall anomalous Hall contribution arises from non-zero Berry curvature of the Bloch states, which leads to an anomalous Hall velocity\cite{chang1996berry, sundaram1999wave, xiao2010berry} when charges are coupled to an electric field.
In the low temperature limit, we show that $d \sigma^{\textrm{int}}_{xy}/d \mu$ is proportional to the average Berry curvature on the Fermi surface and the density of states at the Fermi energy. In particular, this relationship implies a diverging behavior in $d \sigma^{\textrm{int}}_{xy}/d \mu$ as the chemical potential $\mu$ is tuned across the Lishshitz transition, a salient feature that could be observed in low-temperature transport experiments when the anomalous Hall response is dominated by Berry phase effects.

\begin{figure}
    \centering
    \includegraphics[width = 9 cm]{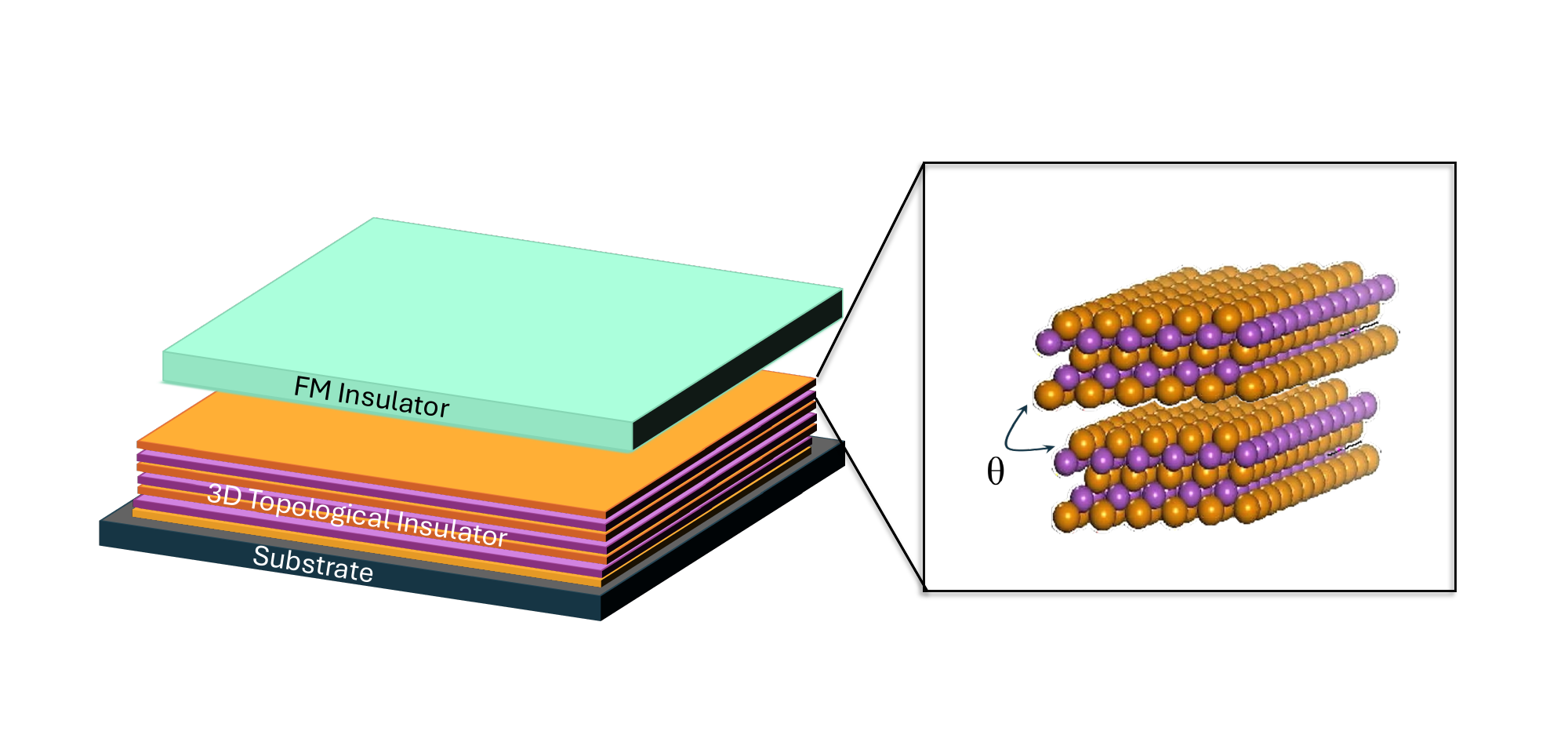}
    \caption{Schematic illustration of the experimental realization of $3$D TI surface with a relative twist between the top quintuple layers, as shown in the inset, in proximity with a ferromagnetic insulator that breaks TRS.}
    \label{fig: Setup}
\end{figure}

This work is organized as follows. In Sec. ~\ref{sec: HOVHS}, we present a model for the realization of HOVHS on moir\'e surface states of $3$D TIs, relying on the interplay of a moir\'e potential and a uniform Zeeman field. We analyze two moir\'e hexagonal superlattices and discuss the effects of inversion symmetry in Sec. ~\ref{subsec: C6}, and of inversion symmetry breaking in Sec. ~\ref{subsec: C3} on the properties of HOHVS. The low energy theory is described by Chern bands with Chern number $\pm 1$ hosting HOVHS on the moir\'e Brillouin zone valleys. In Sec. ~\ref{sec: AHE}, we describe a connection between Lifshitz transitions and the quantum geometric properties of Chern bands, by showing that the divergence in the density of states in Van Hove singularities imprints on the intrinsic quantum anomalous Hall response, and we discuss this effect for the HOVHS Chern bands on the TI surface, as well as other Chern bands supporting logarithmic VHSs. We conclude with a discussion and outlook in Sec. ~\ref{sec: conclusion}.

\section{Higher Order Van Hove Singularities}
\label{sec: HOVHS}

We consider a description of the TI surface states hosting a single Dirac fermion, which is under the effect of a time-reversal invariant periodic superlattice potential $V(\bs{r}) = V(\bs{r} + \bs{R})$ (where $\bs{R}$ is a translation vector), and a time-reversal breaking uniform Zeeman field $h_z$. The single-particle Hamiltonian is 
\begin{subequations}
\begin{equation}
\hat{H} = \int d^{2}\bs{r}\,  \hat{\psi}^{\dagger}(\bs{r}
)
H
\,
\hat{\psi}^{}(\bs{r})
\end{equation}
where $\hat{\psi}^{T}(\bs{r}) = (\psi_{\uparrow}(\bs{r}),\psi_{\downarrow}(\bs{r}))
$ is the two-component spinor with $\psi^{}_{\sigma = \uparrow,\downarrow}(\bs{r})$ 
the electron annihilation operator, and  
\begin{equation}
    \label{eq: model}
    H = v_F\, \hat{z}\cdot(-i\,\bs{\nabla} \times \bs{\sigma}) + V(\bm{r}) \sigma_0 + h_z \sigma_z
    \,,
\end{equation}
\end{subequations}
where $v_F$ is the Fermi velocity,
$\bs{\sigma} = \big(\sigma_x,\sigma_y,\sigma_z \big)$ are  the Pauli matrices, $\sigma_0$ is the identity matrix, and $\hbar =1$. 
The low energy theory depends upon three energy scales: the effective bandwidth $v_F/a$ where $a$ is the superlattice constant, the strength of the moir\'e potential $V$, and the Zeeman energy scale $h_z$. Henceforth, we work in rescaled units where the low-energy physics is determined by the dimensionless parameters $V\,a/v_F$ and $h_z\,a/v_F$.

In TIs such as Bi$_2$Te$_3$ and Bi$_2$Se$_3$, the moir\'e pattern can emerge when the crystals are grown using MBE. For Bi$_2$Te$_3$ with a Fermi velocity of approximately $0.3$ eV nm, \cite{chen2009experimental} a small in-plane rotation of the top layer, which is facilitated by Cu dopants that reduce the interlayer coupling, results in the emergence of moir\'e triangular superlattice with lattice constant $a \approx 13$ nm. \cite{schouteden2016moire} 
In Bi$_2$Se$_3$ with a Fermi velocity of  $0.4$ eV nm \cite{kuroda2010hexagonally} grown using MBE on a substrate, superlattices 
{with periodicity $a$ ranging from $2$ to $7$ nm}
can be constructed from the direct lattice mismatch of the crystal with the substrates like graphene, \cite{song2010topological} FeSe,\cite{wang2012scanning} Au(111),\cite{jeon2011metal} hBN\cite{xu2015van} and {In$_2$Se$_3$. \cite{wang2011superlattices}}

Since the single particle Hamiltonian Eq. \eqref{eq: model} is invariant under translations by a lattice vector $\bs{R}$ of the moir\'e superlattice, the eigenvalue equation for the Bloch states reads
\begin{equation}
\label{eq: Bloch 1}
    H \Psi_{n,\bs{k}} = E_{n,\bs{k}} \Psi_{n,\bs{k}}
    \,,
\end{equation}
where the two-component spinor $\Psi_{n,\bs{k}}(\bs{r}) = e^{i \bs{k} \cdot \bs{r}} U_{n,\bs{k}} (\bs{r})$ with $U_{n,\bs{k}} (\bs{r} + \bs{R}) = U_{n,\bs{k}} (\bs{r})$. From Eq.\eqref{eq: Bloch 1} it follows that
\begin{subequations}
\label{eq: Bloch 2}
\begin{equation}
    \label{eq: eigenK}
    H_{\bs{k}} U_{n,\bs{k}} (\bs{r}) = E_{n,\bs{k}} U_{n,\bs{k}} (\bs{r})
    \,,
\end{equation}
where 
\begin{equation}
{H_{\bs{k}} =  v_F \big[ \hat{z} \cdot (-i \boldsymbol{\nabla} + \bs{k})\times \boldsymbol{\sigma} \big] + V(\bs{r}) \sigma_0  + h_z \sigma_z } 
\,.
\end{equation}
\end{subequations}

We numerically diagonalize Eq.\eqref{eq: Bloch 2} for a total number of 
$242$ bands. 
One of our main findings is that the reconstructed Dirac spectrum obtained from 
Eq.\eqref{eq: Bloch 2} contains a Chern band near charge neutrality supporting cubic HOVHS with power law diverging density of states $\rho(\varepsilon) \sim |\varepsilon|^{-1/3}$, where the energy scale of the HOVHS is set to zero. 

The appearance of HOVHS in this topological band is controlled by $h_z$ and $V$. 
The energy scale $h_z$ corresponds to the Zeeman gap opened in the Dirac spectrum, which can be experimentally realized in heterostructures where the $3$D TI surface is coupled by the proximity effect to a ferromagnetic insulator (FMI). Experimental realizations of the TI-FMI heterostructure of the quintuple layered $3$D TIs such as Bi$_2$Se$_3$ and Bi$_2$Te$_3$, and  FMIs with comparable lattice constants, such as EuS,\cite{wei2013exchange,jiang2016structural} MnSe,\cite{matetskiy2015direct,hirahara2017large} MnTe,\cite{rienks2019large} and Y$_3$Fe$_5$O$_{12}$\cite{lang2014proximity} present evidence for broken TRS on the TI surface. Leveraging on these experimental conditions, we propose a path to realizing moir\'e Chern bands on the TI surface by coupling $3$D TIs where there is an in-plane rotation of the top quintuple layers, with an FMI of comparable lattice constant, as shown in Fig. \ref{fig: Setup}.

The point group symmetries of the moir\'e potential play an important role in the structure of the higher-order Lifshitz transitions. To address the role of point group symmetries, we consider two classes of superlattice potentials with C$_6$ and C$_3$ symmetry.
In the C$_6$ symmetric case, the HOVHS appears as a pair at the valley $\pm \bs{K}$, as shown in Section \ref{subsec: C6}. Breaking the C$_6$ symmetry down to C$_3$ results in a single HOVHS located either at $\bs{K}$ or at $-\bs{K}$ valley, as discussed in Section \ref{subsec: C3}. In what follows, we study each of these cases separately.

\subsection{$C_6$ symmetric moir\'e potential}\label{subsec: C6}

\begin{figure*}
    \centering
    \includegraphics[width = 18 cm]{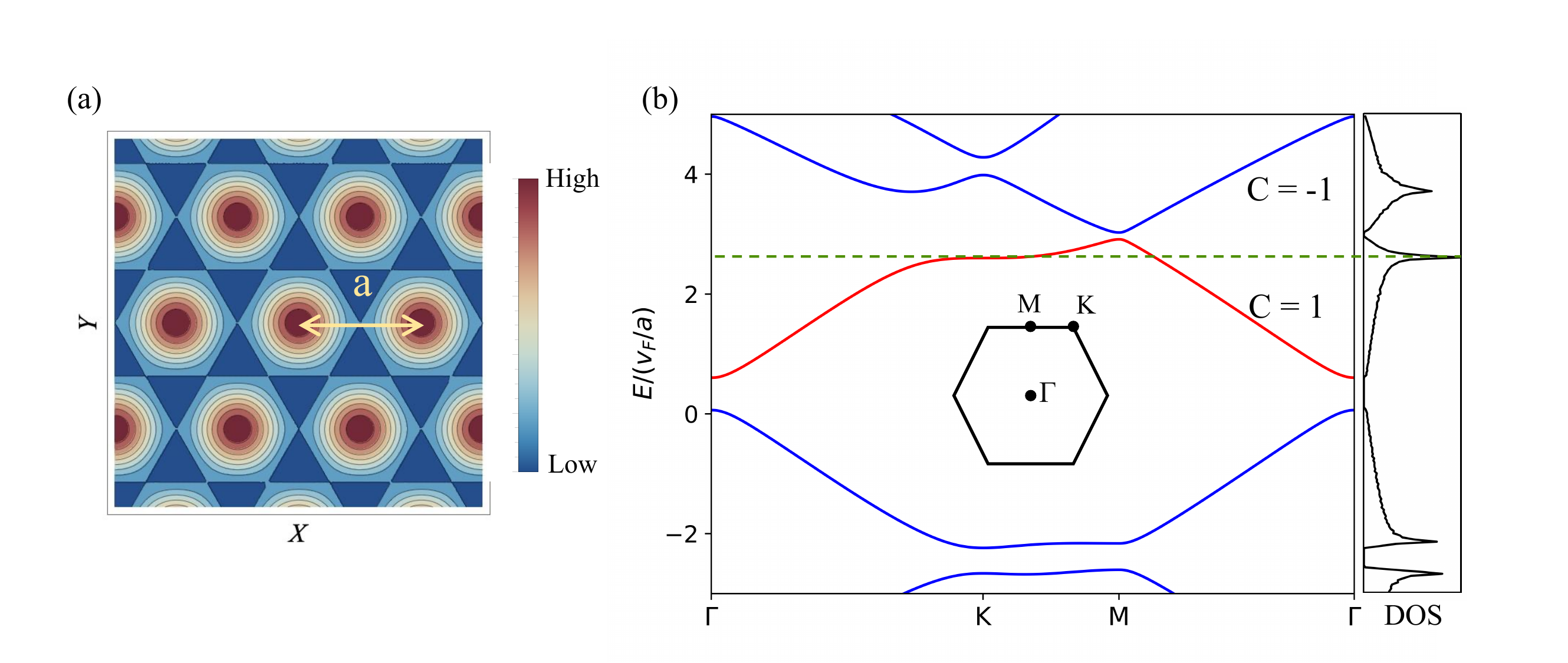}
    \caption{
    (a) Contour plot of the $C_6$ symmetric potential $V_6(\bm{r})$, given in Eq. \eqref{eq: C6pot}, with superlattice constant $a$. 
    (b) Band structure of the model given in Eq. \eqref{eq: model} with the $C_6$ periodic potential for $(h_z, V) = (0.5,1.38)$ in units of $(v_F/a)$ to the left. The mini Brillouin zone is shown in the center. The isolated band shown in red carries the Chern number $C = 1$, and it supports HOVHS at the $\pm \bs{K}$ points, as indicated by the peak in the DOS shown on the right. {The green dashed line denotes the energy at the $\pm \bs{K}$ points, where the DOS diverges.}  
    }
    \label{fig: BandC6}
\end{figure*}

We consider Dirac electrons under the effect of the $C_6$ periodic moir\'e potential as shown in Fig. \ref{fig: BandC6}a
\begin{equation}
    \label{eq: C6pot}
    V_6(\bm{r}) = 2 V \sum_{j=1}^3 \cos(\bm{G}_j \cdot \bm{r})
    \,,
\end{equation}
where $\bm{G}_j = \frac{4 \pi}{\sqrt{3}a} \Big[ -\sin \big(\frac{2 \pi j}{3} \big) , \cos \big(\frac{2 \pi j}{3} \big) \Big]$ are the reciprocal lattice vectors, and $V$ is the strength of the lattice potential. {This potential has invariance under three-fold rotations and inversion symmetry since $V(\bs{r}) = V(-\bs{r})$.}
Without a Zeeman field, the potential given in Eq.\eqref{eq: C6pot} yields a gapless spectrum that supports a band with a pair of HOVHS at the valleys $\pm \bs{K}$ of the moir\'e Brillouin zone when the strength of the potential $V_{0} = 1.36 v_F/a$ \cite{wang2021moire}. Despite this higher-order Lifshtiz transition, these bands remain topologically trivial {due to time-reversal symmetry. At $V = V_0$, the higher-order saddle point at each valley carries a topological index of $-2$ corresponding to the winding of the vector field 
$\bs{\nabla}_{\bs{k}}\varepsilon(\bs{k})$ around the higher-order saddle point.}

To achieve a topological band supporting HOVHS, time-reversal symmetry is broken with the Zeeman field induced by the proximity effect.
On turning the Zeeman field on for $V = V_0$, the higher-order saddle point splits into $4$ critical points, of which $3$ are ordinary saddle points with a topological index of $-1$ each and the remaining one is a local extremum with a topological index of $1$, such that the sum of the topological indices $3 \times (-1) + 1  = 2$ remains conserved.
At the same time, the lowest pair of conduction bands acquire an energy gap and Chern numbers of $\pm 1$ \cite{liu2022magnetic}, as shown in Fig.\ref{fig: BandC6}(b). 

\begin{figure*}
    \centering
    \includegraphics[width = 18 cm]{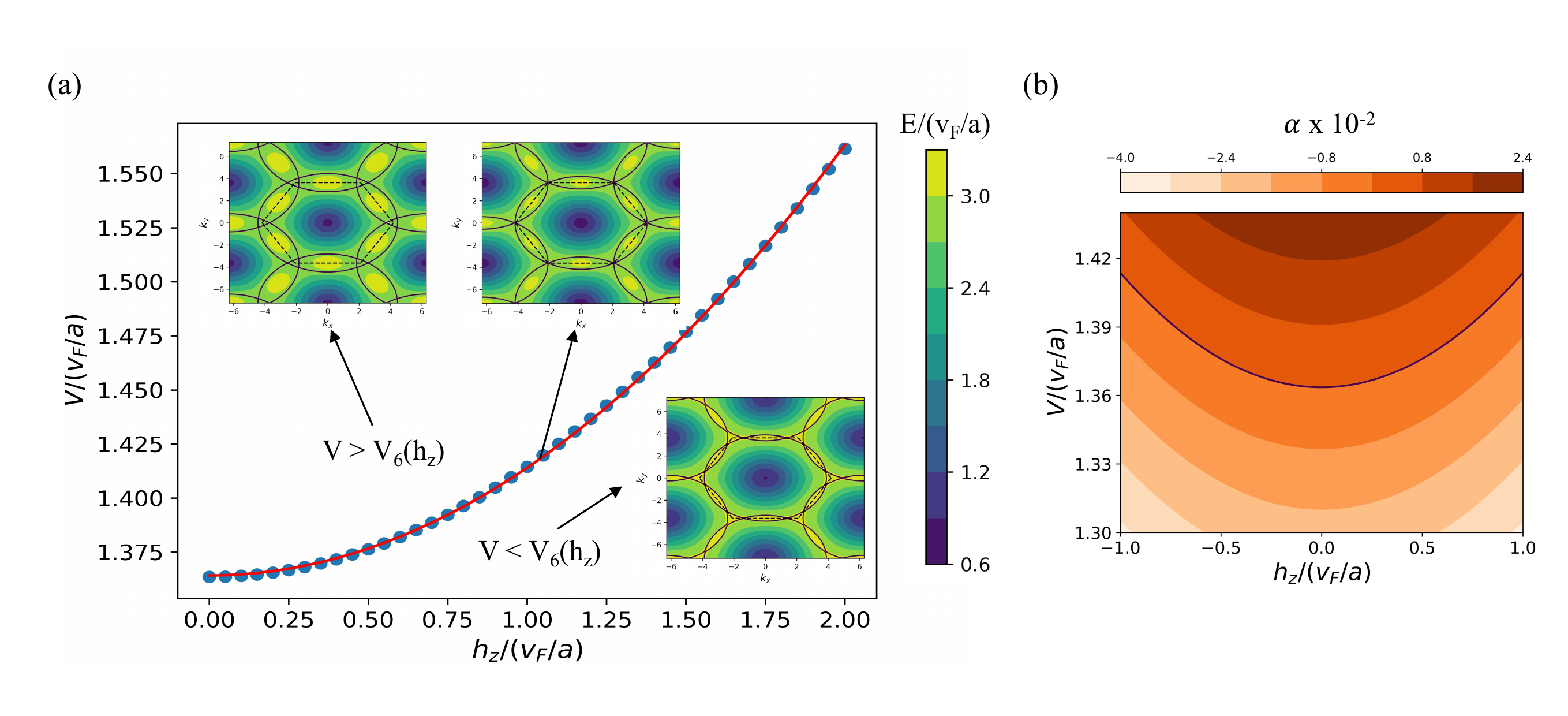}
    \caption{(a) Parameter space plot showing the values of the Zeeman mass ($h_z$) and the lattice potential strength ($V$) that corresponds to HOVHS located at the $\pm \bs{K}$ points of the BZ. The $(h_z, V)$ pairs follow a scaling relation given by Eq.\eqref{eq: hVscaling}. The contour plots of the Fermi surfaces corresponding to the regions- $V>V_6(h_z)$, $V=V_6(h_z)$, and $V<V_6(h_z)$ are shown in the inset, where the dotted black lines indicate the boundaries of the FBZ and the black lines show the Fermi surfaces passing through VHS.
    (b) $\alpha(h_z, V)$ obtained using Eq. \eqref{eq: alpha_C6} plotted in the $h_z-V_6$ parameter space. The thick black curve indicates the $\alpha(h_z, V) = 0$ line which, as expected, coincides with the $h_z-V$ curve shown in (a) that corresponds to the emergence of HOVHS at $\pm \bs{K}$ points.}
    \label{fig: C6alpha}
\end{figure*}

Remarkably, we notice that these split critical points (i.e., three conventional VHS and one extremum) can be merged together again at $\pm \bs{K}$ to generate a HOVHS in the Chern band marked in red in Fig.\ref{fig: BandC6}(b) by adjusting either $V$ or $h_z$. We uncover a line in $(h_z, V)$ parameter space, shown in Fig. \ref{fig: C6alpha}(a), where the $C=1$ band supports pairs of HOVHS. This line in $(h_z, V)$ parameter space obeys the relation
\begin{equation}
    \label{eq: hVscaling}
    V_6(h_z) = V_6^{(0)} + \Gamma\, h_z^2
    \,,
\end{equation}
where, $V_6^{(0)} = V_6(h_z=0) = 1.36 v_F/a$, and $\Gamma \approx 0.05$ is a fitting parameter. 
{The quadratic scaling between $h_z$ and $V$ is observed for $h_z$ values ranging from $0$ to $\approx 2v_F/a$. As a result, for moir\'e surface states with lattice constant $a \approx 10$ nm and Fermi velocity $v_F \approx 300 $ meV. nm, the parameters $h_z$ and $V$ that can be employed to tune the HOVHS follow the scaling relation given in Eq. \eqref{eq: hVscaling} for $h_z$ as large as $60$ meV}

To understand the relation given by Eq. \eqref{eq: hVscaling}, we study the energy dispersion around the valley $\bs{K}$, 
\begin{equation}
    \label{eq: EK_C6}
    \varepsilon(\bs{p}) \equiv E(\bm{p} + \bm{K}) - E(\bm{K}) = \alpha p^2 + \beta (p_x^3 - 3 p_x p_y^2)
    \,,
\end{equation}
where $\bm{p} = (p_x,p_y)$ is the momentum in the vicinity of the valley. The momentum dependence of the dispersion Eq. \eqref{eq: EK_C6}, expanded up to the third order, is dictated by the point group symmetries of the hexagonal lattice. From this, 
we obtain the critical points $\bs{\nabla}_{\bs{p}}\varepsilon(\bs{p}) = 0$ and characterize their behavior by computing the Hessian $\mathcal{H} = \mathrm{det}\Big( \partial_{p_i} \partial_{p_j} \varepsilon(\bs{p})\Big)$.

While the gradient of the dispersion vanishes for $\bm{p} \in \Big\{(0,0),\big(\frac{-2 \alpha}{3 \beta},0 \big), \big(\frac{ \alpha}{3 \beta},\pm \frac{\alpha}{\sqrt{3} \beta} \big) \Big\}$, the corresponding Hessian $\mathcal{H} = 4(\alpha^2 - 9 \beta^2 p^2)$ evaluated at each of the four aforementioned critical points 
vanishes only when the coefficient $\alpha = 0$, which implies the vanishing of the term quadratic in momentum.
Thus, the scaling relation Eq. \eqref{eq: hVscaling} characterizes the points in the $(h_z, V)$ parameter space for which the quadratic in momentum coefficient $\alpha(h_z,V) = 0$, so that the energy dispersion around $\bs{K}$ can be described by a third-order polynomial $\varepsilon(\bs{p}) \approx \beta (p_x^3 - 3 p_x p_y^2)$, corresponding to  a HOVHS with diverging density of states $\rho(\varepsilon) \sim \frac{1}{|\varepsilon|^{1/3}}$.~\cite{shtyk2017electrons}

Furthermore, in the range of $(h_z,V)$ values where we observe the quadratic scaling given in Eq. \eqref{eq: hVscaling}, the parameter $\alpha(h_z,V)$ can be well approximated by the empirical form,
\begin{equation}
    \label{eq: alpha_C6}
    \alpha(h_z,V) = \alpha_0 \ln \Big[ \frac{V}{ V_6(h_z)}\Big]
    \,,
\end{equation}
where $\alpha_0 \approx 0.4$, and $V_6(h_z)$ is given by Eq. \eqref{eq: hVscaling}. {The manifold of $(h_z, V)$ values for which $\alpha(h_z, V)$ vanishes is denoted by the black curve in Fig. \ref{fig: C6alpha}(b).}
When the strengths of the potential and of the Zeeman energy deviate from condition \eqref{eq: hVscaling}, the HOVHS splits into three ordinary saddle point VHS for $\bm{p} \in \Big\{\big(\frac{-2 \alpha}{3 \beta},0 \big), \big(\frac{ \alpha}{3 \beta},\pm \frac{\alpha}{\sqrt{3} \beta} \big) \Big\}$ and a local extremum at $\bm{p} = (0,0)$. 
{Under small deviations $\delta V$ of the moir\'e potential and the Zeeman energy from $(h_z = 0, V = V_{6})$, a perturbative expansion of the coefficient $\alpha(h_z, \delta V)$ to lowest order yields $\alpha(h_z, V) \approx c_1\,\delta V + c_2\, h^{2}_{z}$, where a liner term in $h_z$ vanishes due to inversion symmetry. Then the condition $\alpha(h_z, \delta V) = 0$ yields Eq. \eqref{eq: hVscaling}.}

\begin{figure}
    \centering
    \includegraphics[width = 8.45 cm]{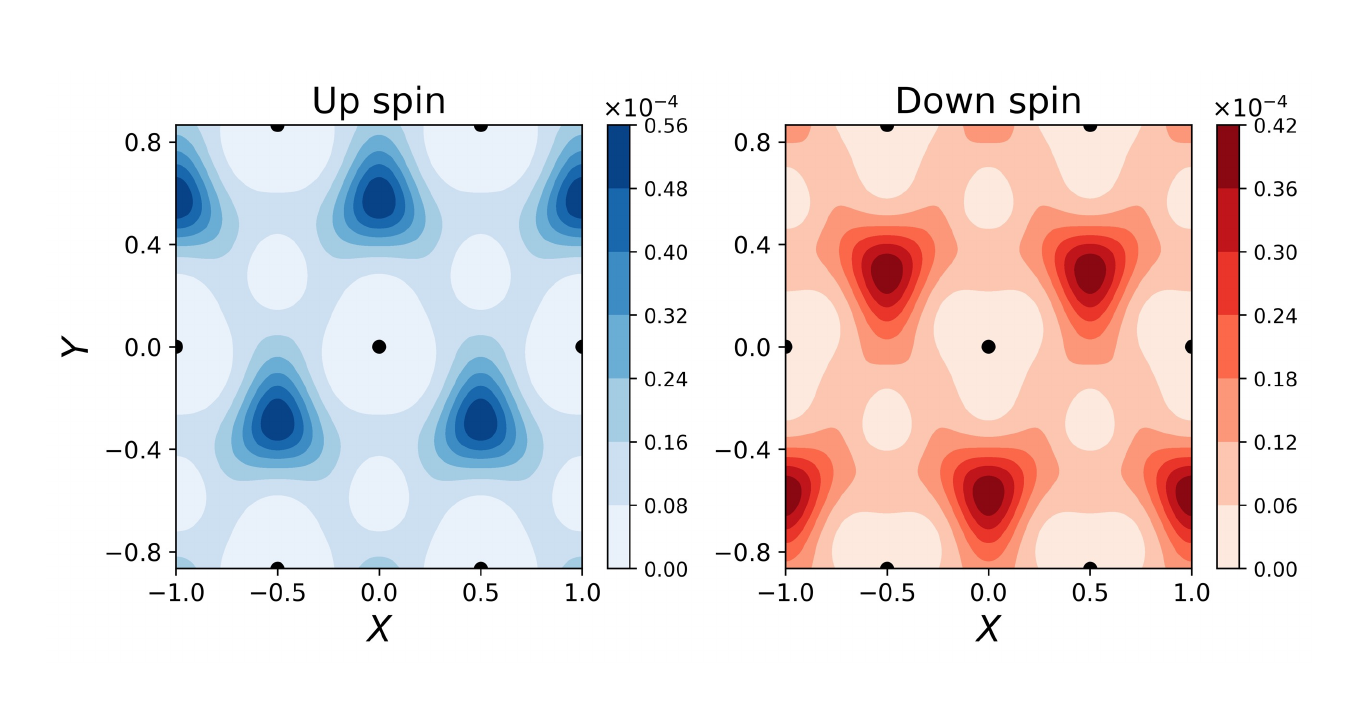}
    \caption{ Real space wavefunction of the up (left) and down (right) spin electrons with momentum $\bs{K}$ on the HOVHS band {of the $C_6$ symmetric model with} $(h_z, V) = (0.5,1.38)$ in units of $(v_F/a)$. The difference in the maximum amplitude of the up and down spins can be attributed to the broken TRS. The maxima of the lattice potential are denoted with black dots. }
    \label{fig: Pol}
\end{figure}

Apart from the high concentration of density of states, the bands supporting HOVHS display another interesting feature. As shown in Fig. \ref{fig: Pol}, the real space wave functions for the energy states at the HOVHS and projected on the valleys $\pm \bs{K}$ are particularly polarized on two sublattices that form a honeycomb lattice localized between the maxima of the moir\'e potential. This sublattice polarization behavior is observed along the HOVHS line given by Eq. \eqref{eq: hVscaling}, and highlighted in Fig. \ref{fig: Pol} for $(h_z, V) = (0.5,1.38)$.
We note that this sublattice polarization in the vicinity of HOVHS on the TI surface states is a feature observed in other two-dimensional systems tight-binding systems such as in Haldane Chern insulators tuned to higher-order Lifshitz transitions \cite{castro2023emergence} and kagom\'e bands. \cite{kiesel2012sublattice, wu2023sublattice}

\subsection{$C_3$ symmetric moir\'e potential}\label{subsec: C3}

\begin{figure*}
    \centering
    \includegraphics[width = 18 cm]{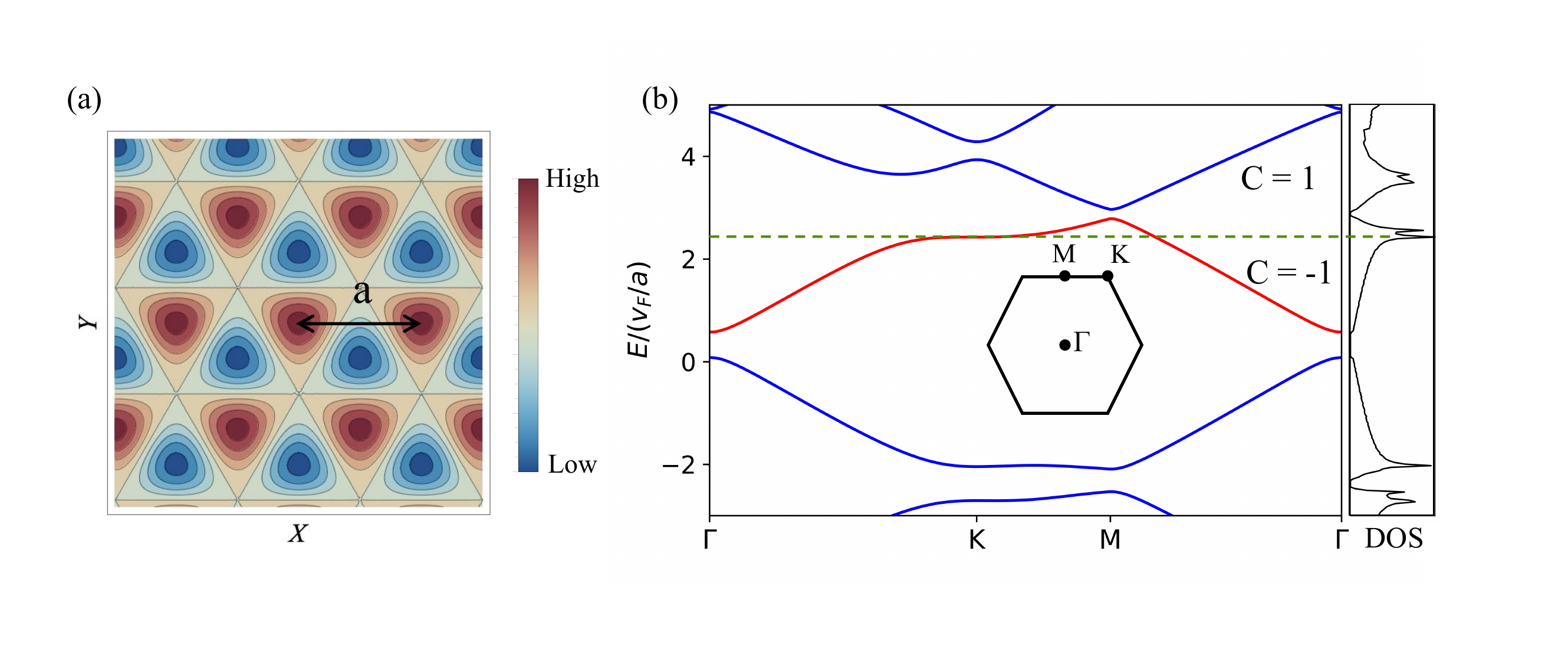}
    \caption{
    (a) Contour plot of the $C_3$ symmetric potential $V_3(\bm{r})$, given in Eq. \eqref{eq: C3pot}, with superlattice constant $a$. 
    (b) Band structure of the model given in Eq. \eqref{eq: model} with the $C_3$ periodic potential with $(h_z, V) = (0.5,0.83)$ in units of $(v_F/a)$, to the left and the corresponding DOS to the right. The sharp peak in the DOS at the $\bs{K}$ point indicates the presence of HOVHS while the shorter peak right above it in energy points towards the conventional VHS around the $\bs{K^{\prime}}$ point. Here, the lowest conduction band (shown in red) that carries a Chern number of $-1$, hosts HOVHS located at the $\bs{K}$ points.
    }
    \label{fig: BandC3}
\end{figure*}

We investigate the effects of breaking inversion symmetry by adding a contribution to the $C_6$ symmetric lattice potential given in Eq. \eqref{eq: C6pot} that yields a potential with $C_3$ symmetry 
\begin{equation}
    \label{eq: C3pot}
    V_3(\bm{r}) = 2 V \sum_{j=1}^3 \Big[ \cos(\bm{G}_j \cdot \bm{r}) - \cos(\bm{G}_j \cdot \bm{r} + \phi) \Big] 
    \,,
\end{equation}
where $\phi = \frac{2 \pi}{3}$, as depicted in Fig. \ref{fig: BandC3}a. 
In the absence of the Zeeman field, this model supports a gapless spectrum of moir\'e bands, wherein we uncover one band supporting a pair of HOVHS a the moir\'e Brillouin zone valleys when the strength of the moir\'e potential $V_3^{(0)} = V_3(h_z=0) = 0.802 (v_F/a)$. The existence of a pair of valley HOVHS follows from TRS degenerate states at $\pm\bs{K}$ despite breaking of inversion symmetry by the potential Eq. \eqref{eq: C3pot}.

On turning the Zeeman field on with $V = V_3^{(0)}$, the HOVHS splits into 3 conventional VHS located around each of the $\bs{K}$ and $-\bs{K}$ points of the lowest conduction band which carries a Chern number of $-1$, as shown in Fig. \ref{fig: BandC3}b. 
By adjusting the lattice potential strength or the Zeeman field, the conventional VHS can merge back into a HOVHS. However, due to the breaking of inversion symmetry, this HOVHS is located in either one of the valleys. This situation is analogous to the effect of an inversion-breaking sublattice potential on the Haldane Chern insulator model that can lead to a single HOVHS on one of the valleys. \cite{aksoy2023single}

\begin{figure*}
    \centering
    \includegraphics[width = 17 cm]{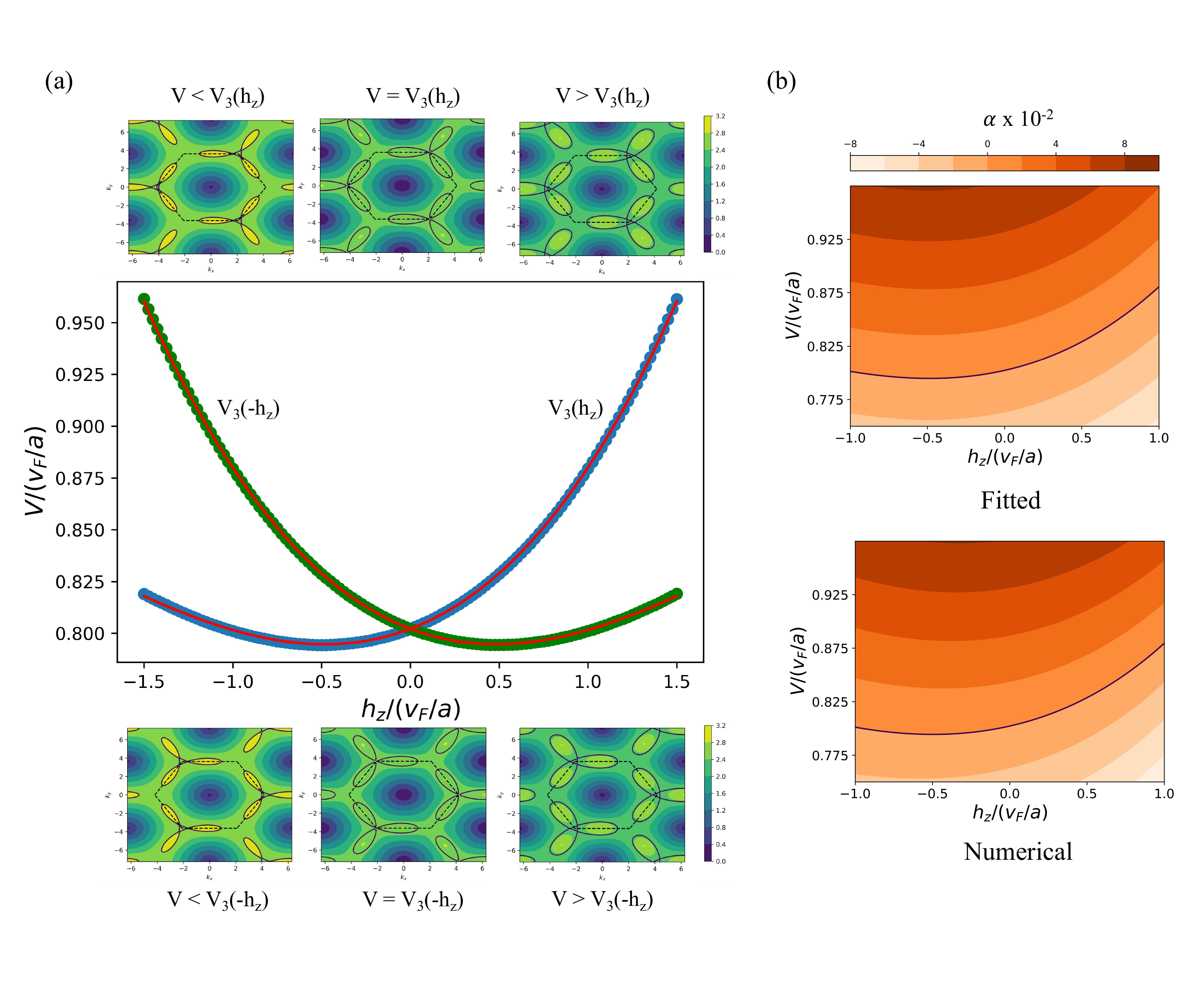}
    \caption{
    (a)Parameter space plot showing the values of the Zeeman mass ($h_z$) and the lattice potential strength ($V$) that corresponds to HOVHS located at the $\bs{K}$ (blue) and the $\bs{K^{\prime}}$ (green) points of the BZ. The $(h_z, V)$ pairs that support HOVHS located at K ($K^{\prime}$) follow a scaling relation $V_3(h_z)$ ($V_3(-h_z)$) given by Eq. \eqref{eq: hVscalingC3}. The contour plots of the Fermi surfaces corresponding to the regions- $V<V_3(h_z)$, $V=V_3(h_z \neq 0)$, and $V>V_3(h_z)$ are shown above, and those corresponding to the regions- $V<V_3(-h_z)$, $V=V_3(-h_z \neq 0)$, and $V>V_3(-h_z)$ are shown below. When $h_z=0$ and $V = V_3^{(0)}$, we find HOVHS located at both $\bs{K}$ and $\bs{K^{\prime}}$ points, as in the case of the $C_6$ symmetric lattice potential. Here, the dotted black lines indicate the boundaries of the first moir\'e Brillouin zone and the black lines show the Fermi surfaces passing through VHS.
    (b)$\alpha(h_z, V)$ obtained numerically from the expression given in Eq. \eqref{eq: alphaC3} (top) and from the Taylor expansion of the energy dispersion around the $\bs{K}$ points (bottom). The thick black curve indicates the $\alpha(h_z, V) = 0$ line which, as expected, coincides with the $h_z-V$ curve corresponding to the emergence of HOVHS at the $\bs{K}$ point shown in (a).
    }
    \label{fig: C3alpha-hV}
\end{figure*}

{Similarly to the approach of Sec. \ref{subsec: C6}, we
characterize the $(h_z, V)$ parameter space for which the $C=-1$ band supports HOVHS in one of the valleys.
For the $\bs{K}$ valley HOVHS, we uncover the relation between moir\'e potential and Zeeman field,
\begin{equation}
    \label{eq: hVscalingC3}
    V_3(h_z) = V_3^{(0)} + \Gamma_1 h_z + \Gamma_2 h_z^2 + \Gamma_3 h_z^3
    \,,
\end{equation}
in terms of numerically fitted coefficients $(\Gamma_1, \Gamma_2, \Gamma_3) \approx (0.033,0.039,0.007)$, and $h_z$ and $V${are measured in units of $(v_F/a)$}. We note that unlike in the case discussed in Sec. \ref{subsec: C6}, the breaking of inversion symmetry allows for terms with odd powers in $h_z$. Furthermore, while $(h_z, V_3(h_z))$ corresponds to a higher-order saddle point at the $\bs{K}$ point, $(h_z, V_3(-h_z))$ corresponds to the one at the $-\bs{K}$ point. In Fig. \ref{fig: C3alpha-hV}(a), we plot these curves and the character of the Fermi surface contours at and in the vicinity of the higher-order Lifshitz transitions. 
}

Eq. \eqref{eq: hVscalingC3}, which describes the condition for the coefficient $\alpha(h_z, V)$ in Eq. \eqref{eq: EK_C6} to vanish, can be well approximated by 
\begin{equation}
    \label{eq: alphaC3}
    \alpha(h_z,V) = \alpha_0 \ln \Big[ \frac{V}{ V_3(h_z)}\Big] 
    \,,
\end{equation}
where $\alpha_0 \approx 0.4$ for $h_z a/v_F < 2$. {Similarly, the Hessian vanishes at the $-\bs{K}$ points when $\alpha(h_z,V) = \alpha_0 \ln (V/V_3(-h_z))$ with $\alpha_0 \approx 0.4$.}
Thus, we have established a mechanism to create and access a larger landscape of HOVHS topological bands with Chern number $C = \pm 1$, which occur through the interplay of a moir\'e potential and a uniform Zeeman field coupled to the surface states a 3D topological insulator. In the next section, we discuss an important feature in anomalous Hall response implied by the existence of Van Hove singularities in Chern bands and apply it to the case of conventional and higher-order VHS.

\section{Intrinsic anomalous Hall conductivity near Lifshitz transitions} \label{sec: AHE}

{In Chern bands, the presence of non-zero Berry curvature of the Bloch states leads to an anomalous Hall velocity\cite{chang1996berry, sundaram1999wave, xiao2010berry} when charges couples to an electric field, giving rise to the intrinsic anomalous Hall effect.\cite{karplus1954hall,jungwirth2002anomalous,nagaosa2010anomalous}}
While for a filled Chern band insulator, this leads to a quantized Hall conductivity in units of $e^2/h$, in a partially filled Chern band, the intrinsic anomalous Hall conductivity $\sigma^{\textrm{int}}_{xy}$ can be continuously tuned by the electronic filling of the Chern band, and it probes the Berry phase contribution of the occupied electronic states. In particular, when the Fermi energy crosses a Lifshitz transition in a Chern band, the large peak in the density of states entailed by the change in the topology of the Fermi surface suggests that the anomalous Hall response may correspondingly manifest some distinct property.
Furthermore, this distinct behavior should be a generic feature of Chern bands, provided the sharp features in the density of states are not significantly rounded by disorder and thermal broadening. 
Thus, we hereafter discuss the character of the intrinsic anomalous Hall response of Chern bands at low temperatures and in cases where the anomalous Hall effect is dominated by the intrinsic contribution\cite{nagaosa2010anomalous}, where we uncover a characteristic feature relating the anomalous Hall conductivity and the density of states. Specifically, in the $T \rightarrow 0$ limit, we show that $d\,\sigma^{\textrm{int}}_{xy}/d\,\mu$ diverges when the Fermi energy $\mu$ crosses the scale of logarithmic and power-law Van Hove singularities. Also, in what follows, we assume that the system remains a Fermi liquid as the Fermi energy crosses a Van Hove singularity.

To establish these connections, we consider an isolated Chern band described by Bloch states $\ket{u_{\bs{k}}}$ with energy dispersion $\varepsilon(\bs{k})$ and non-vanishing Berry curvature $\Omega(\bs{k}) = \nabla_{\bs{k}} \cross \bs{A}_{\bs{k}}$, where the Berry connection $\bs{A}_{\bs{k}} = i\,\bra{u_{\bs{k}}}\nabla_{\bs{k}}\,\ket{u_{\bs{k}}}$. This system is characterized by the intrinsic anomalous Hall conductivity \cite{karplus1954hall}
\begin{equation}
\label{eq: sigma xy 1}
\sigma^{\textrm{int}}_{xy}(\mu;T) = \frac{e^2} {h}\, \frac{1}{2 \pi}\, \int\,d^{2} \bs{k}\,  \Omega(\bs{k}) f_{\mu;T}(\varepsilon(\bs{k})) 
\,,
\end{equation}
where the integral extends over the first Brillouin zone, $e$ is the charge of the electron, $h$ is Planck's constant and
$
f_{\mu;T}(\varepsilon(\bs{k}))  = \Big({e^{\frac{\varepsilon(\bs{k})-\mu}{k_{B}T}}+1}\Big)^{-1}
$
is the Fermi distribution at temperature $T$ and chemical potential $\mu$.  

In the $T \rightarrow 0$ limit, Eq. \eqref{eq: sigma xy 1} becomes 
\begin{equation}
\label{eq: sigma xy 2}
\sigma_{xy}^{\textrm{int}}(\mu;0) = \frac{e^2} {h}\, \frac{1}{2 \pi}\,\int\,d^{2}\bs{k}\,  \Omega(\bs{k}) \Theta(\mu-\varepsilon(\bs{k}))
\,,
\end{equation}
where $\Theta(x)$ is the Heaviside step function signaling a sharp separation between occupied and empty electronic states. This contribution varies continually with the Fermi energy $\mu$ when the Fermi energy lies between the minimum and maximum of the Chern band.

An insight can be obtained by considering differential anomalous Hall conductivity, which measures the slope of $\sigma_{xy}^{\textrm{int}}$ as a function of Fermi energy,
\begin{equation}
\label{eq: d sigma at T=0}
\frac{ d \sigma^{\textrm{int}}_{xy}(\mu;0)}{ d \mu}= \frac{e^2} {h}\, \frac{1}{2 \pi}\,\int\,d^{2}\bs{k}\,  \Omega(\bs{k}) \delta(\mu-\varepsilon(\bs{k}))
\,.
\end{equation}
Eq. \eqref{eq: d sigma at T=0} shows that at $T=0$, the differential anomalous Hall conductivity gets its contribution from states at the one-dimensional Fermi surface manifold $\varepsilon(\bs{k}) = \mu$. 

\begin{widetext}
\begin{figure*}
\begin{center}
\includegraphics[width = 17 cm]{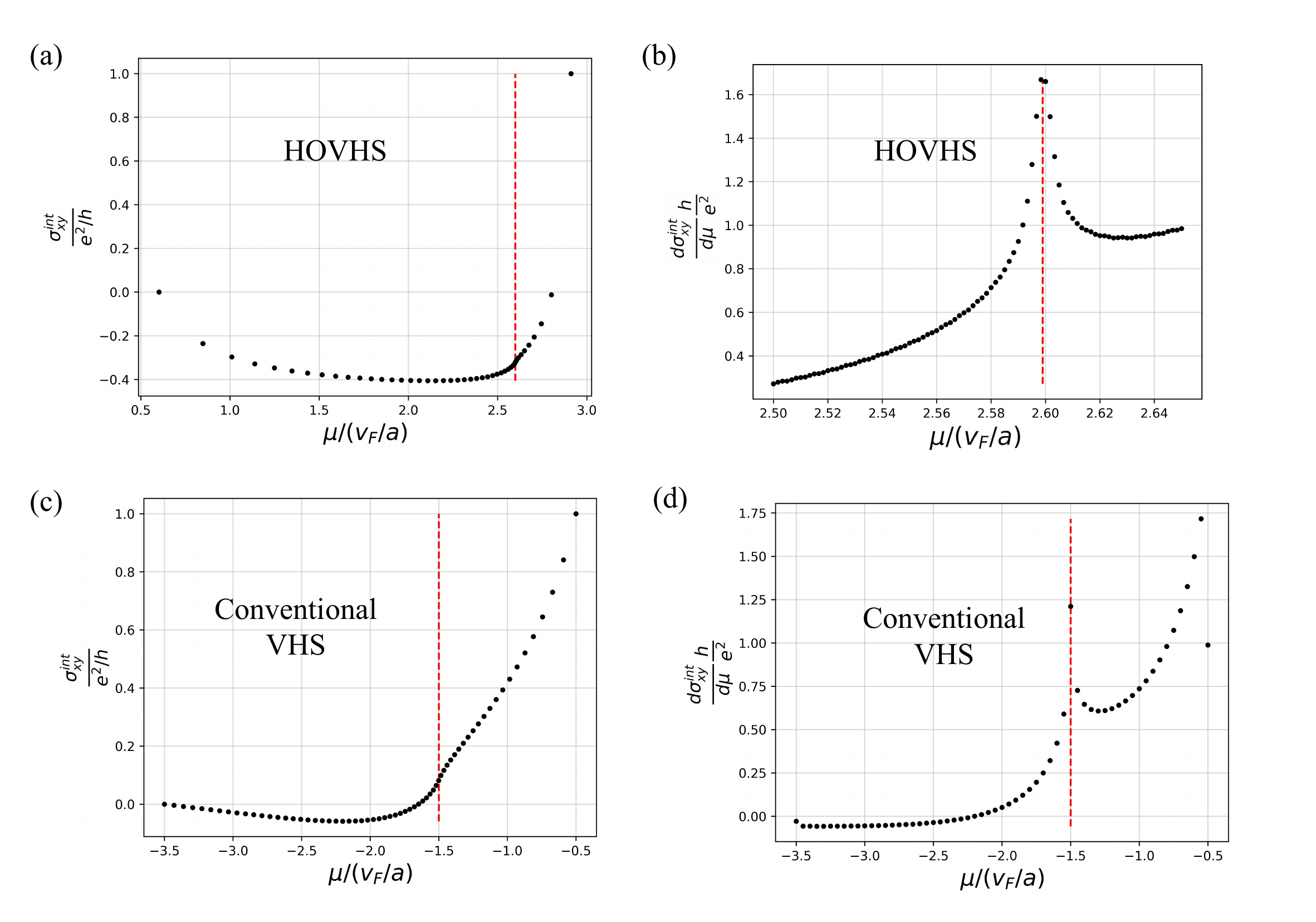}
\end{center}
\caption{(a) Intrinsic anomalous Hall conductivity ($\sigma^{\textrm{int}}_{xy}$) and (b) differential anomalous Hall response ($\frac{d\sigma^{\textrm{int}}_{xy}}{d \mu}$), at zero temperature plotted as a function of the Fermi energy $\mu$ for the Chern band supporting HOVHS at $(h_z, V) = (0.5,1.38)$ in units of $(v_F/a)$. The dashed red line denotes the energy at which HOVHS occurs, $\mu_{*}$. The differential anomalous Hall response exhibits a power law divergence around $\mu_{*}$, as given in Eq. \eqref{eq: general relation} (b), { with $\nu = 0.39 \approx 1/3$, and $\kappa_{+} = \kappa_{-} = 0.43$}.
(c) Intrinsic anomalous Hall conductivity ($\sigma^{\textrm{int}}_{xy}$) and (d) differential anomalous Hall response ($\frac{d\sigma^{\textrm{int}}_{xy}}{d \mu}$), at zero temperature plotted as a function of the Fermi energy $\mu$ for the lower energy band of the Chern insulator model defined on a square lattice of lattice constant $a$, $H = (v_F/a)[\sin k_x  \sigma_x + \sin k_y  \sigma_y + (m- \cos k_x  - \cos k_y ) \sigma_z]$ at $m = 1.5$ where it supports VHS at $\mu_* = -1.5 (v_F/a)$ (denoted by the red dashed line) and carries a Chern number of $1$. The differential anomalous Hall response exhibits a log divergence around $\mu_{*}$, as given in Eq. \eqref{eq: general relation} (a), with $\Lambda = 1.35 (v_F/a) = 0.45 \times$ bandwidth and $\rho_0 = \frac{0.58}{v_F/a}$.
Note that in (a) and (c), as the Fermi reaches the top of the band, the total Hall conductivity approaches $1$, the Chern number of the corresponding band.
}
\label{fig: sigma}
\end{figure*}
\end{widetext}

Upon normalizing by the density of states at the Fermi energy,
\begin{equation}
\label{eq: DOS}
\rho(\mu) =  \int\,d^{2}\bs{k}\, \delta(\mu-\varepsilon(\bs{k})) 
\,,
\end{equation}
Eq.  \eqref{eq: d sigma at T=0} 
reads
\begin{equation}    
\label{eq: general relation sigma and DOS}
\frac{ d \sigma^{\textrm{int}}_{xy}(\mu;0)}{ d \mu}= \frac{e^2} {2 \pi h}\, \langle \Omega \rangle_{FS} \, \rho(\mu)
\,,
\end{equation}
where
\begin{equation}
\langle \Omega \rangle(\mu)
\equiv
\frac{\int\,d^{2}\bs{k}\,  \Omega(\bs{k}) \delta(\mu-\varepsilon(\bs{k}))}
{\int\,d^{2}\bs{k}\,\delta(\mu-\varepsilon(\bs{k}))}  
\end{equation}
defines the average of the Berry curvature on the Fermi surface. 

Eq. \eqref{eq: general relation sigma and DOS} provides an insightful connection between the average Berry curvature (which is the imaginary part of the quantum metric tensor), the density of states, and the anomalous Hall transport, which can be in principle explored to read out the Berry curvature by scanning the Fermi energy of the band.
Moreover, let us consider the behavior of Eq. \eqref{eq: general relation sigma and DOS} in the vicinity of a Lifshitz transition occurring at energy $\mu^{*}$. In this case, if $\langle \Omega \rangle(\mu^{*}) \neq 0$ (which we expect to occur typically in a Chern band), the differential anomalous Hall conductivity is dominated by the Van Hove singularities in the density of states, resulting in the following asymptotic forms for the conventional and higher-order Lifshitz transitions 
\begin{widetext}
\begin{subequations}
\label{eq: general relation}
\begin{equation}
\begin{split}
\frac{ d \sigma_{xy}(\mu;0)}{ d \mu}
&\,
\approx
\frac{e^2} {2 \pi h}
\,\times
\langle \Omega \rangle (\mu)
\,\times
\underbrace{\rho_{0}\,\log(\Lambda/|\mu - \mu_{*}|)}_{\textrm{Conventional Lifshitz transition}}
\,,
\end{split}
\end{equation}  
\begin{equation}
\begin{split}
\frac{ d \sigma_{xy}(\mu;0)}{ d \mu}
&\,
\approx
\frac{e^2} {2 \pi h}
\,\times
\langle \Omega \rangle (\mu)
\,\times\,\,
\underbrace{\frac{{\Theta(\mu-\mu_{*})\kappa_{+} + \Theta(\mu_{*}-\mu)\kappa_{-} }}{|\mu - \mu_{*}|^{\nu}} }_{\textrm{Higher-order Lifshitz transition}}
\,,
\end{split}
\end{equation}
\end{subequations}
\end{widetext}
where $\nu$ is a positive exponent and $\kappa_{\pm}$ coefficients for $\mu \geq \mu_{*}$ ($\mu \leq \mu_{*}$).
Eq.~\eqref{eq: general relation} establishes a general connection between differential anomalous Hall response and Lifshitz transitions in Chern bands.

In Figure \ref{fig: sigma}, we plot $d \sigma^{\textrm{int}}_{xy} / d\mu$ 
and $\sigma^{\textrm{int}}(\mu)$ as a function of the Fermi energy $\mu$ for two representative cases, namely, the Chern bands holding HOVHS discussed in Sec.\ref{sec: HOVHS} and a two-band model Chern insulator model supporting logarithmic VHS \cite{qi2006general,qi2008topological}. In the top panels of Figure \ref{fig: sigma}, we observe that the cubic power-law divergence in the DOS for the topological insulator moir\'e surface Chern bands gives rise to a pronounced peak in $d \sigma^{\textrm{int}}_{xy} / d\mu$ (Figure \ref{fig: sigma}b), which then reflects a characteristic slope in $\sigma^{\textrm{int}}_{xy}$ (Figure \ref{fig: sigma}a) as the system undergoes a Lifshitz transition at $\mu = \mu^{*}$, which is indicated by the red dotted lines. In the bottom panels of Fig. \ref{fig: sigma}, we perform a similar analysis for a Chern band with conventional logarithmic VHS. In both cases, the characteristic behavior by Eq. \eqref{eq: general relation} is confirmed despite some numerical rounding introduced by the finite-size grid.

\begin{figure}
    \centering
    \includegraphics[width = 9.15 cm]{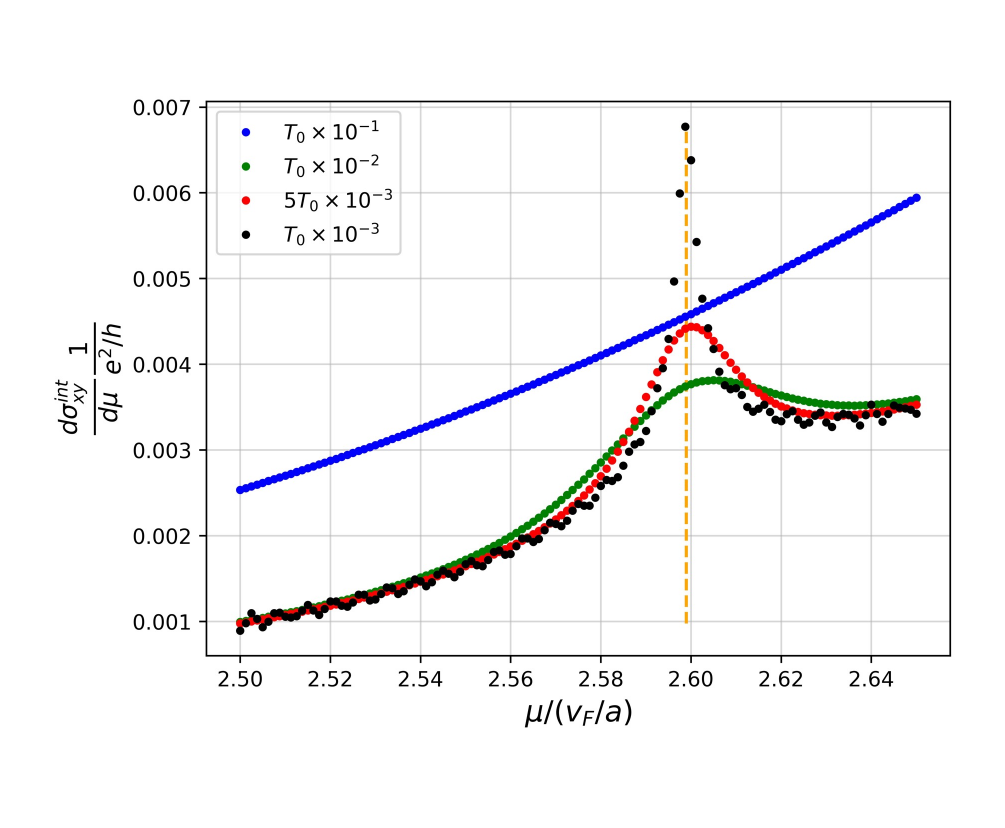}
    \caption{Differential anomalous Hall conductivity $\frac{d\sigma^{\textrm{int}}_{xy}}{d \mu}$ plotted against the Fermi energy $\mu$ for the HOVHS band with $(h_z, V) = (0.5,1.38)$ in units of $(v_F/a)$, for different values of temperature ($T$) that are labelled on the plot. The temperature $T$ is scaled with $T_0 = (v_f/a)/k_B$ where $k_B$ is the Boltzmann constant and $v_F/a \approx 30$ meV. \cite{schouteden2016moire, chen2009experimental}
    The yellow dashed line denotes the energy at which the HOVHS occurs, $\mu_{*}$. The peak in $\frac{d\sigma^{\textrm{int}}_{xy}}{d \mu}$ associated with the HOVHS gets progressively stronger as the temperature decreases.}
    \label{fig: dsigmaT}
\end{figure}

To account for the effects of thermal broadening, we obtain from Eq. \eqref{eq: sigma xy 1} the temperature dependence of the differential anomalous Hall conductivity
\begin{equation}
    \label{eq: dSigmaTdef}
    \begin{split}
&\, \frac{d\, \sigma^{\textrm{int}}_{xy}}{d\, \mu} (\mu; T) = 
\\
&\,
\frac{e^2}{2 \pi\, h}\, \int\, d^2\, \bs{k}\,\Omega(\bs{k})\,\frac{1}{k_B\,T}\Big[ 2\, \cosh{\frac{(\varepsilon(\bs{k}) - \mu)}{2\,k_B\,T}} \Big]^{-2}
    \,,
    \end{split}
\end{equation}
where the last term in the integral accounts for the temperature dependence of the Fermi distribution.

In Figure \ref{fig: dsigmaT} we plot the temperature dependence of the differential anomalous Hall conductivity Eq. \eqref{eq: dSigmaTdef} for the TI surface state discussed in Sec. \ref{subsec: C6} for the Chern band supporting a pair of HOVHS. 
As expected, for temperatures compared with or greater than the bandwidth $T \gtrsim T_{0} \sim \Lambda/k_{B}$, where $\Lambda$ is the bandwidth, thermal effects strongly destroy the effect of the VHS. However, with decreasing of temperature substantially below the bandwidth scale, the peak in $\frac{d\, \sigma^{\textrm{int}}_{xy}}{d\, \mu}$ becomes progressively more pronounced, allowing for the identification of a VHS, and asymptotically tending to a sharp peak as in $T \rightarrow 0$ limit described by Eq.~\eqref{eq: general relation}.

\section{Discussion and Outlook} \label{sec: conclusion}

In this work, we have identified a mechanism to create time-reversal broken topological Chern bands that host higher-order Van Hove singularities on the surface of a 3D topological insulator. We have shown that these HOVHS in topologically nontrivial bands can emerge from the interplay of a time-reversal breaking Zeeman field induced by the proximity to a ferromagnetic insulator and a time-reversal invariant moir\'e potential induced on the surface electrons of a three-dimensional topological insulator. The latter can naturally occur through the misalignment of the quintuple layers in Bi$_2$Se$_3$ and Bi$_2$Te$_3$, responsible for the onset of a nanometer scale moir\'e potential. Employing exact diagonalization and symmetry analysis, we have demonstrated that tuning of the Zeeman and moir\'e potential energy scales gives rise to a manifold of higher order Lifshitz transitions on the moir\'e Brillouin zone valleys. 
This setting opens a direction for future exploration of correlation effects associated with the presence of strong density of state singularities in moir\'e Chern bands on the surface of topological insulators.

Furthermore, we have identified a characteristic signature in the intrinsic anomalous Hall response as the Fermi surface crosses a Lifshitz transition. Specifically, the rate of change of the anomalous Hall conductivity as a function of the Fermi energy,
$d\sigma^{\textrm{int}}_{xy}/ d \mu$, displays a pronounced peak at low temperatures due to the large accumulation of states near the VHS, which tracks the
logarithmic or power-law divergences in conventional and higher-order Lifshitz transitions, respectively. In the moir\'e surface Chern bands studied in this work, this entails a power-law divergence in $d\sigma^{\textrm{int}}_{xy}/ d \mu$ as the temperature tends to zero. This relationship opens a route to experimentally probe VHS and Chern bands through transport measurements, and it would be not only applicable in the HOVHS 
on the surface of a topological insulator but also to a wider class of Hofstadter-Chern bands,
\cite{wang_classification_2020,Herzog-Arbeitman_Hofstadter2020,Spanton2018,sharpe2019emergent,serlin2020intrinsic,Saito2021,xie2021fractional,nuckolls2020strongly,wu_chern_2021,das2021symmetry,choi2021correlation,park2021flavour,stepanov2021competing,pierce2021unconventional,yu2021correlated}
and zero-field Chern bands in moir\'e heterostructures.\cite{cai2023signatures, park2023observation, zeng2023thermodynamic, xu2023observation, lu2024fractional}

\section*{Acknowledgments}
We thank Ben Feldman for the valuable discussions. This research was supported by the U.S. Department of Energy, Office of Science, Basic Energy
Sciences, under Award DE-SC0023327. L. P. acknowledges funding from the Women in Natural Science Fellowship of Emory University.

\bibliographystyle{apsrev4-1} 
\bibliography{ref}

\end{document}